\documentclass[twocolumn,aip,apl,reprint, superscriptaddress,preprintnumbers,amsmath,amssymb]{revtex4-1}

\usepackage[pdftex]{graphicx}
\usepackage{epstopdf}
\usepackage{graphicx}
\usepackage{dcolumn}
\usepackage{bm}
\usepackage{amssymb}
\usepackage{wasysym}
\usepackage{amsmath} 
\usepackage{amsthm}
\usepackage{bibtopic}
\usepackage{sidecap}
\usepackage[rflt]{floatflt}
\usepackage{float}

\begin{document}

\def\Ef{$E_{\rm F}$}
\def\Eb{$E_{\rm B}$}
\def\Efmath{E_{\rm F}}
\def\Ed{$E_{\rm D}$}
\def\Tc{$T_{\rm C}$}
\def\kpara{{\bf k}$_\parallel$}
\def\kparamath{{\bf k}_\parallel}
\def\kperp{{\bf k}$_\perp$}
\def\Gbar{$\overline{\Gamma}$}
\def\Kbar{$\overline{K}$}
\def\Mbar{$\overline{M}$}
\def\BiTe{Bi$_2$Te$_3$}
\def\BiSe{Bi$_2$Se$_3$}
\def\SbTe{Sb$_2$Te$_3$}
\def\Ed{$E_{\rm D}$}
\def\invA{\AA$^{-1}$}

\title{Laser-induced persistent photovoltage on the surface of a ternary topological insulator at room temperature}

\author{J. S\'anchez-Barriga}
\email[Corresponding author. E-mail address: ] {jaime.sanchez-barriga@helmholtz-berlin.de.}
\affiliation{Helmholtz-Zentrum Berlin f\"ur Materialien und Energie, Albert-Einstein-Str. 15, 12489 Berlin, Germany}
\author{M. Battiato}
\affiliation{Institute of Solid State Physics, Vienna University of Technology,  Vienna A-1040, Austria} 
\author{E. Golias}
\affiliation{Helmholtz-Zentrum Berlin f\"ur Materialien und Energie, Albert-Einstein-Str. 15, 12489 Berlin, Germany}
\author{A. Varykhalov}
\affiliation{Helmholtz-Zentrum Berlin f\"ur Materialien und Energie, Albert-Einstein-Str. 15, 12489 Berlin, Germany}
\author{L. V. Yashina}
\affiliation{Department of Chemistry, Moscow State University, Leninskie Gory 1/3, 119991, Moscow, Russia}
\author{O. Kornilov}
\affiliation{Max-Born-Institut, Max-Born-Str. 2A, 12489 Berlin, Germany}
\author{O. Rader}
\affiliation{Helmholtz-Zentrum Berlin f\"ur Materialien und Energie, Albert-Einstein-Str. 15, 12489 Berlin, Germany}

\begin{abstract}
Using time- and angle-resolved photoemission, we investigate the ultrafast response of excited electrons in the ternary topological insulator (Bi$_{1-x}$Sb$_{x}$)$_2$Te$_3$ to fs-infrared pulses. We demonstrate that at the critical concentration $x$=0.55, where the system becomes bulk insulating, a surface voltage can be driven at room temperature through the topological surface state solely by optical means. We further show that such a photovoltage persists over a time scale that exceeds $\sim$6 $\mu$s, i.e, much longer than the characteristic relaxation times of bulk states. We attribute the origin of the photovoltage to a laser-induced band-bending effect which emerges near the surface region on ultrafast time scales. The photovoltage is also accompanied by a remarkable increase in the relaxation times of excited states as compared to undoped topological insulators. Our findings are relevant in the context of applications of topological surface states in future optical devices.
\end{abstract}

\maketitle

Topological insulators \cite{Hasan-RMP-2010} (TIs) are an exotic state of matter with unique properties that can be utilized in future optoelectronic applications. Their insulating bulk property is due to an inverted band structure arising from high spin-orbit interaction,\cite{Zhang-NatPhys-2009} and their metallic surface is characterized by Dirac cone-like topological surface states (TSSs) with high spin-polarization,\cite{Yazyev-PRL-2010,Hsieh-Nature-2009,Souma-PRL-2011,Pan-PRL-2011,Jozwiak-PRB-2011,Sanchez-Barriga-PRX-2014} spin-momentum locking,\cite{Hsieh-Science-2009,Roushan-Nature-2009} and protection by time-reversal symmetry.\cite{Fu-PRL-2009} These properties render TSSs very promising for the exploitation of spin-dependent transport in devices.

However, one of the major challenges for the utilization of TIs in future information technology is the suppression of their residual bulk conductance, which overwhelms the surface contribution.\cite{Ando-JPSoc-2013} This situation is usually found in prototypical TIs such as Bi$_2$Se$_3$, Bi$_2$Te$_3$, and Sb$_2$Te$_3$, where the Fermi level does not lie within the bulk band gap. \cite{Zhang-NatPhys-2009,Ando-JPSoc-2013} Therefore, a tremendous effort has been devoted to develop TI materials that exhibit bulk-insulating behavior in transport, as this is of critical importance to take advantage of the unprecedented properties of TSSs in actual devices.\cite{Ando-JPSoc-2013} Recently, an intrinsic surface transport regime has been achieved in a variety of complex TI compounds.\cite{Ando-JPSoc-2013,Ren-PRB-2011,Arakane-NatComm-2012,Kuroda-PRB-2015,Kushwaha-PRB-2015} One of the most prominent examples is the ternary TI (Bi$_{1-x}$Sb$_{x}$)$_2$Te$_3$,\cite{JZhang-NatComm-2011} in which proper tuning of the concentration $x$ allowed the observation of the surface quantum Hall effect.\cite{Yoshimi-NatComm-2015} It was also shown that this system exhibits the quantum anomalous Hall effect when additional magnetic dopants are introduced into the bulk.\cite{Chang-Sience-2013}

In parallel with these material efforts, the use of fs-laser pulses has offered an alternative route to solely excite surface carriers within the bulk band gap of a TI on ultrafast time scales.\cite{Gedik-2013-Science-Floquet,Sobota-PRL-2012-Bulk-Reservoir,Kuroda-PRL-2016} This approach is promising for the generation and optical manipulation of surface spin currents that are relevant for spintronics.\cite{Pesin-NatMat-2012} In this context, light-matter interaction plays the most important role towards the utilization of topological properties in future optical devices. In particular, it has been shown that exciting a TI with fs-laser pulses allows the creation of a surface photovoltage (SPV).\cite{Neupane-PRL-2015} However, up to date, a SPV effect in TIs has been observed only in Bi$_2$Te$_2$Se, which was obtained with a defect density allowing to achieve the intrinsic surface transport regime. \cite{Neupane-PRL-2015,Ren-PRB-2010} 

Therefore, it is of critical importance to understand whether the emergence of a SPV effect in TIs is material-specific or a general phenomenon that also exists in other TIs that are in the bulk-insulating regime. Moreover, for TI-based device applications, it is essential to provide compelling evidence that the SPV effect persists at room temperature. Equally important is the understanding of the underlying mechanism that in TIs governs the relaxation behavior of the SPV on ultrafast time scales. It should be noted that a SPV has been also observed in various semiconductors such as GaAs or Si,\cite{Tokudomi-JPSoc-2007,Ogawa-PRB-2013} however, these materials are topologically trivial and the interplay between the SPV and topological properties is fundamentally absent. 

To fully address these issues, systematically investigating the same TI compound by tuning its composition in order to access the whole doping range is necessary. Moreover, in the context of TI-based applications, for example in optical p-n junctions, sensors, or all-optical gating, the possibility to fabricate these components by precisely tuning the doping in selected regions within the same TI compound is more desirable, as it would facilitate the exploitation of the SPV effect in potential devices that are exclusively based on TIs.

Hence, in the present work we perform time-resolved angle-resolved photoemission (tr-ARPES) measurements on the ternary TI (Bi$_{1-x}$Sb$_{x}$)$_2$Te$_3$ as a function of composition following optical excitation by fs-infrared pulses. We observe a persistent SPV at the critical concentration $x$=0.55 when the system is in the bulk-insulating regime. We demonstrate that the SPV can be optically driven at room temperature through the TSS, which exhibits a relaxation time scale more than two orders of magnitude larger than that in the bulk-conducting regime. We further discuss the origin of the SPV in terms of a laser-induced band-bending effect emerging on ultrafast time scales due to a charge accumulation process near the surface region. 

Tr-ARPES experiments were performed under ultrahigh vacuum conditions with a base pressure below $1\times10^{-10}$ mbar. Emitted photoelectrons were detected at room temperature with a Scienta R4000 hemispherical analyzer using linearly-polarized pump (1.5 eV) and probe (6 eV) pulses incident on the sample under an angle of $\phi=45^{\circ}$. (Bi$_{1-x}$Sb$_{x}$)$_2$Te$_3$ single crystals were grown by the Bridgman method using different compositions from the melt and cleaved {\it in situ}. The solidification time of the crystals was $\sim$1 week, and the pulling rate $\sim$0.2 mm/h. The temperature gradient during growth was 5$^{\circ}$C/cm and growth temperatures were in the range of 580--600$^{\circ}$C. The infrared-pump and ultraviolet-probe pulses were generated with a homemade fs-laser system coupled to an ultrafast amplifier operating at 150 kHz repetition rate. The time resolution of the experiment was $\sim$200 fs, and the pump fluence $\sim$100 $\mu$J/cm$^{2}$. The angular and energy resolutions of the tr-ARPES measurements were 0.3$^{\circ}$ and 20 meV, respectively. 

Figures~1(a)-1(d) display energy-momentum band dispersions of (Bi$_{1-x}$Sb$_{x}$)$_2$Te$_3$ measured under equilibrium conditions for few selected Sb concentrations $x$ that are representative for the evolution of the band structure. We clearly observe a progressive upward shift of the TSS and bulk-valence band (BVB) states that is consistent with a transformation of the system from $n$- to $p$-type with increasing $x$. In Figs.~1(e)-1(h), we zoom-in on the region near the Fermi level of Figs.~1(a)-1(d), respectively. The data are represented using a logarithmic intensity scale to better visualize the contribution from the less intense bulk-conduction band (BCB) states in this energy region. The corresponding energy-distribution curves (EDCs) at zero momentum (\kpara=0) are also shown on the right side of each panel (red circles). 

The parent compounds Bi$_2$Te$_3$ ($x$=0) and Sb$_2$Te$_3$ ($x$=1) possess high degrees of metallicity originating from the bulk. Bi$_2$Te$_3$ [Figs.~1(a) and 1(e)] exhibits $n$-type doping with the Fermi level located deep inside the BCB, while Sb$_2$Te$_3$ [Figs.~1(d) and 1(h)] is naturally $p$-doped with the Fermi level in the BVB. Thus, the TSS bands in Sb$_2$Te$_3$ are largely unoccupied in equilibrium. For $x$=0.38 [Figs.~1(b) and 1(f)], the BCB becomes less intense because it moves upwards by about 90 meV as compared to Bi$_2$Te$_3$. Despite the reduced intensity, the BCB bottom can be observed below the Fermi level in the energy-momentum dispersion of Fig.~1(f), leading to a peak in the corresponding EDC at \kpara=0. Therefore, for x=0.38 the Fermi level is outside of the bulk band gap, and the main difference with respect to Bi$_2$Te$_3$ is that the number of available BCB states below the Fermi level substantially decreases. On the other hand, at the critical concentration $x$=0.55 [Figs.~1(c) and 1(g)], the contribution from BCB states at the Fermi level completely disappears, meaning that the sample is in a bulk-insulating state with the Fermi level inside the bulk band gap. This result is consistent with previous observations of the antipolar field effect in nanoplates as derived from bulk single crystals of the same critical composition.\cite{Kong-NatNanotech-2011} Moreover, note that in Figs.~1(c) and 1(g) only the two linear branches of the TSS bands that do not hybridize with the bulk above the Dirac point are crossing the Fermi level. This type of crossing is especially important because from energies below and up to the Dirac point the surface bands directly overlap with the BVB.

\begin{figure}
\centering
\includegraphics [width=0.4\textwidth]{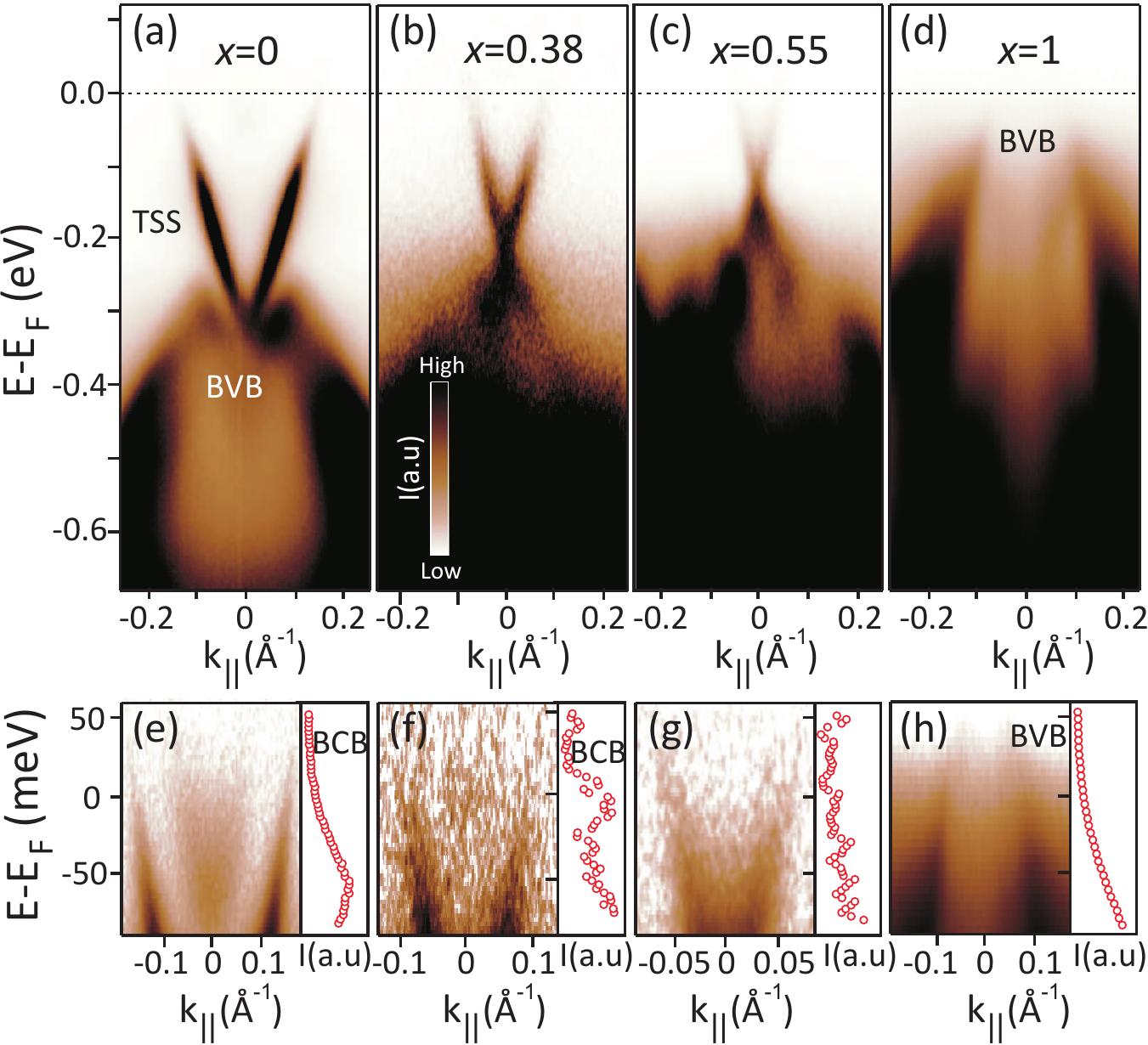}
\caption{(a)-(d) Selected energy-momentum band dispersions of (Bi$_{1-x}$Sb$_{x}$)$_2$Te$_3$ obtained in equilibrium along the \Gbar-\Kbar\ direction of the surface Brillouin zone as a function of composition $x$. (e)-(h) Zoom-in on the region near the Fermi level of (a)-(d), respectively. The data are represented using a logarithmic intensity scale to fully visualize the contributions from bulk states in this region. On the right side of each panel, the corresponding EDCs at zero momentum are shown (red circles).}
\label{Fig1}
\end{figure}

To understand the nonequilibrium dynamics of excited states as a function of doping $x$, we performed systematic tr-ARPES measurements in the whole composition range. In Fig. 2, to immediately visualize the impact that suppressing the bulk metallicity has on the dynamics, we compare selected tr-ARPES spectra at various pump-probe delays for $x$=0.55 [Fig.~2(a)-2(e)] and $x$=1 [Fig.~2(f)-2(j)]. Very clearly, in both cases we observe an initial excitation of the TSS and BCB up to $\sim$0.8 eV above Fermi level [Figs.~2(b) and 2(g)]. In the subsequent dynamics, the higher energy states rapidly relax within few ps [Figs.~2(c) and 2(i)]. After this initial relaxation, both systems exhibit a transient chemical potential near the bottom of the BCB, which for $x$=1 decays much faster [Fig.~2(j)]. Instead, very interestingly, for x=0.55 the transient chemical potential survives more than 200 ps and stabilizes within the TSS once all electrons from the BCB have completely decayed. 
\begin{figure}
\centering
\includegraphics [width=0.35\textwidth]{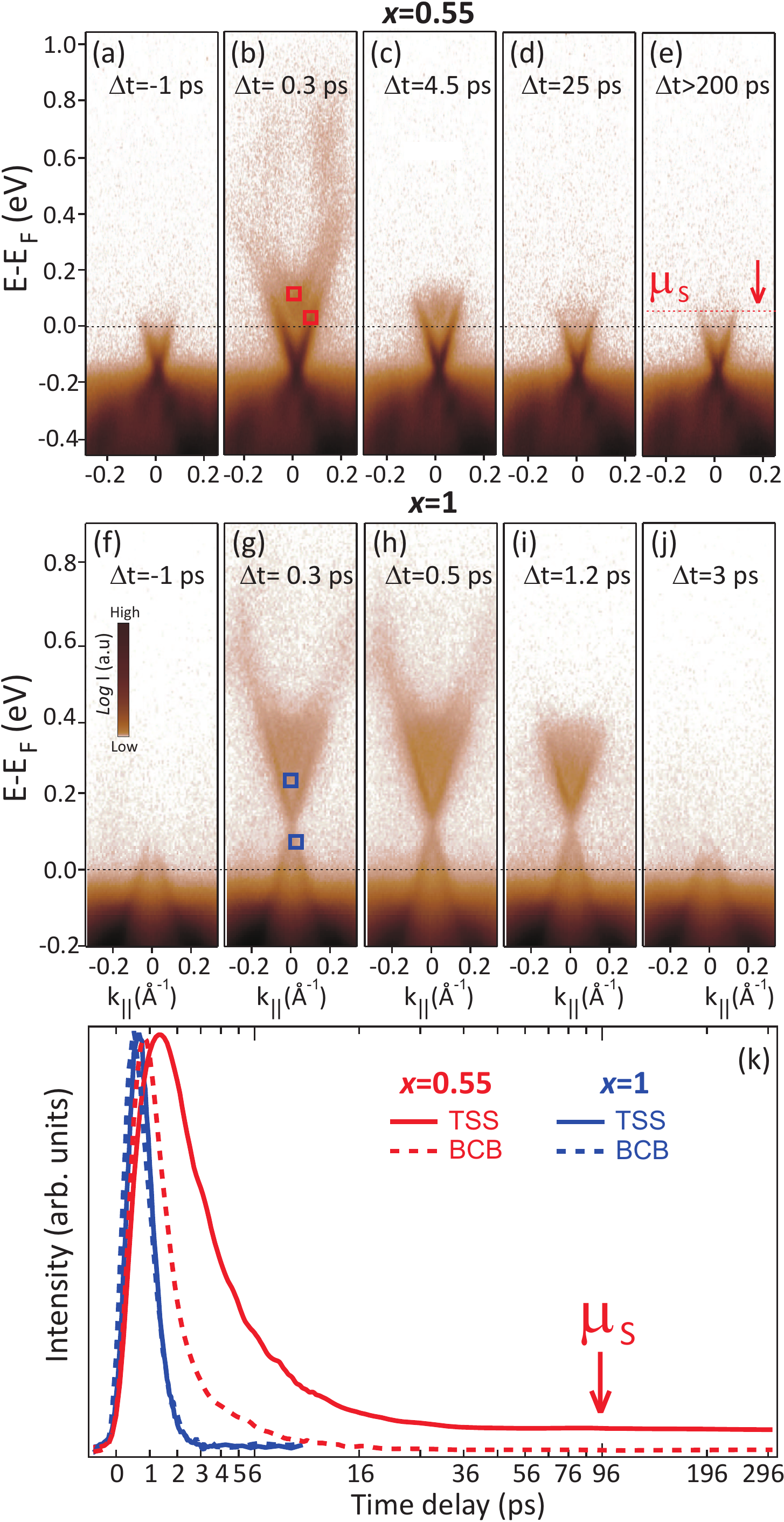}
\caption{Comparison between tr-ARPES spectra for (a)-(e) $x$=0.55 and (f)-(j) $x$=1 obtained along \Gbar-\Kbar\ at various pump-probe delays, as indicated on the top of each panel. (k) Tr-ARPES intensities of bulk (dashed lines) and surface states (solid lines) for $x$=0.55 (red) and $x$=1 (blue) as a function of pump-probe delay. The intensities are integrated within the small energy-momentum windows shown in (b) and (f), respectively. The contribution from a long-lived transient chemical potential $\mu_S$ observed in (e) is also indicated in (k).}
\label{Fig2}
\end{figure}

\begin{figure}
\centering
\includegraphics [width=0.35\textwidth]{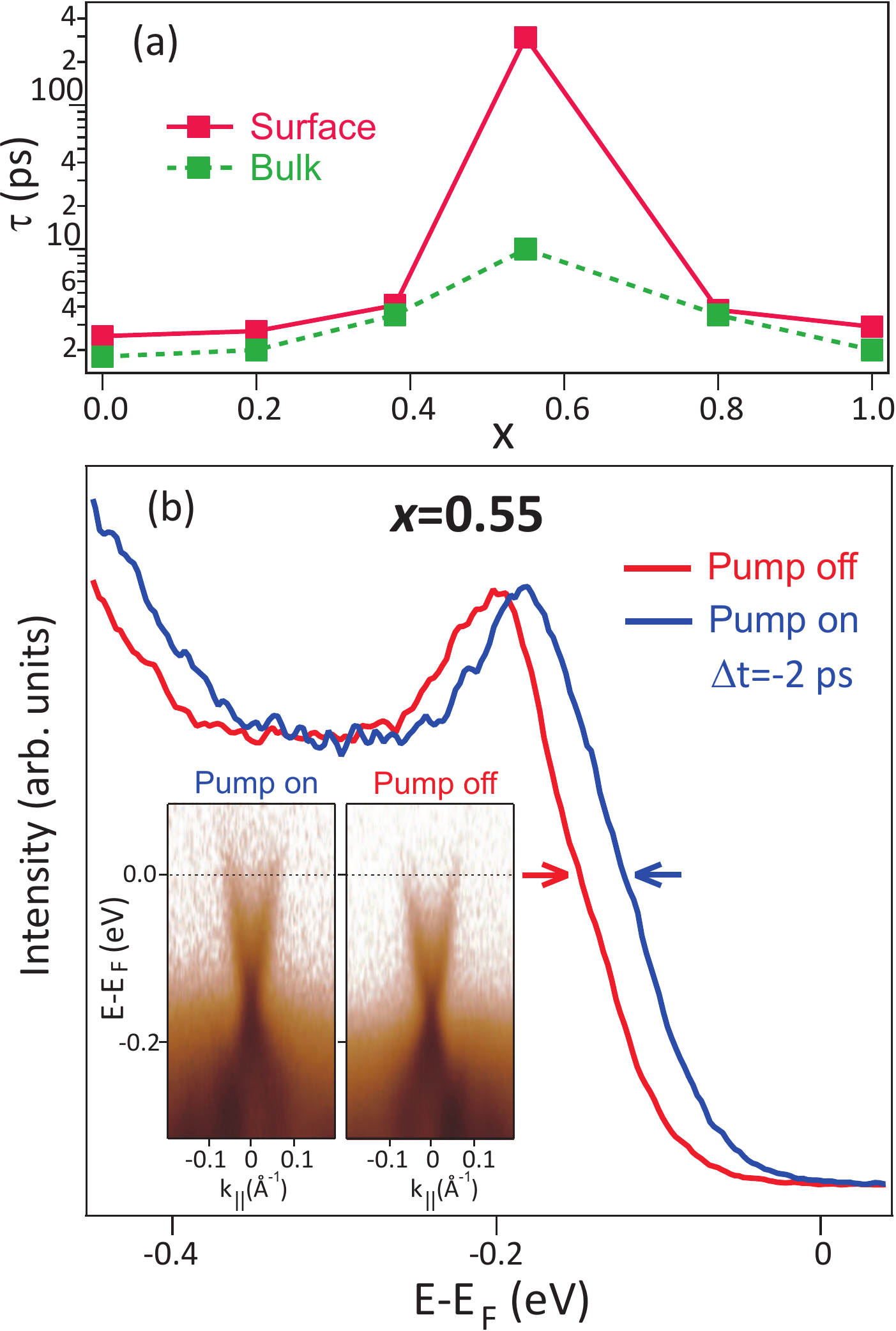}
\caption{(a) Relaxation times of surface (red) and bulk states (green) in (Bi$_{1-x}$Sb$_{x}$)$_2$Te$_3$ as a function of concentration $x$. (b)
A quasi-static contribution to the SPV is clearly observed as an energy shift in the photoemission signal without (red) and with laser pump at $\Delta$t=-2 ps (blue) for $x$=0.55. The spectra are EDCs extracted at zero momentum from the corresponding energy-momentum dispersions shown as an inset.}
\label{Fig3}
\end{figure}
At these long time delays, the transient chemical potential [denoted as $\mu_s$ and indicated by a red dashed line in Fig.~2(e)] amounts to $\sim$70 meV when compared to the position of the Fermi level [black dashed lines in Fig.~2]. The long-lived contribution from $\mu_s$ can be cleary seen in Fig.~2(k), where we plot the time dependence of the integrated intensity within the small energy-momentum boxes shown in Figs.~2(b) and 2(g). The boxes are located above the Fermi energy and at the lowest representative energy of the BCB and TSS bands. The size of the boxes was chosen to be similar to the energy and momentum broadening of the TSS bands in order to integrate as much intensity as possible from each individual band. For $x$=1 both bulk and surface bands have comparable decay times of $\sim$2 ps (blue dashed and solid lines, respectively). Instead, for $x$=0.55, electrons within the BCB decay at much smaller rate and their dynamics persits up to $\sim$10 ps (red dashed line). Moroever, TSS electrons decay at nearly five times slower rate (red solid line), and their dynamics converges into a metastable state that is fully governed by the transient chemical potential $\mu_s$. Such a metastable state persists up to pump-probe delays outside of the measured time window [see, e.g., Fig.~2(a)].

In Fig.~3(a) we compare the relaxation time ($\tau$) of the bulk (green) and surface (red) populations for the different Sb concentrations $x$ analyzed in the present work. The remarkable increase observed only for $x$=0.55 indicates that this effect is not material-specific but rather a general phenomenon for TIs that are in the bulk-insulating regime. We note that a noticeable but much smaller increase in $\tau$ is observed for $x$=0.38, a result that we attribute to the reduced number of available bulk states in the immediate vicinity of the Fermi level as compared to $x$=0 and $x$=1. On the other hand, for $x$=0.55, the lack of available bulk states at the Fermi level strongly suppresses the phase space for electron-hole recombination. In this respect, the difference of about two orders of magnitude in the relaxation times of surface and bulk electrons for this composition strongly suggests that there is a charge accumulation process near the surface region that acts as a bottleneck for the dynamics as long as electrons on the surface cannot effectively diffuse into the bulk. Therefore, it is plausible to conclude that the long-lived transient chemical potential observed in Fig.~2 is directly associated with an emergent SPV in the bulk-insulating regime. It should be emphasized that within this scenario, the mechanism underlying the decay of the excited population would be solely related to a charge transport phenomenon. In other words, at extremely long time scales we would expect an increase of the potential gradient in the near surface region due to a spatial delocalization of the surface charge into the bulk. In this case, we would expect a measurable SPV due to a laser-induced band-bending effect emerging near the surface region on ultrafast time scales. 

\begin{figure}
\centering
\includegraphics [width=0.3\textwidth]{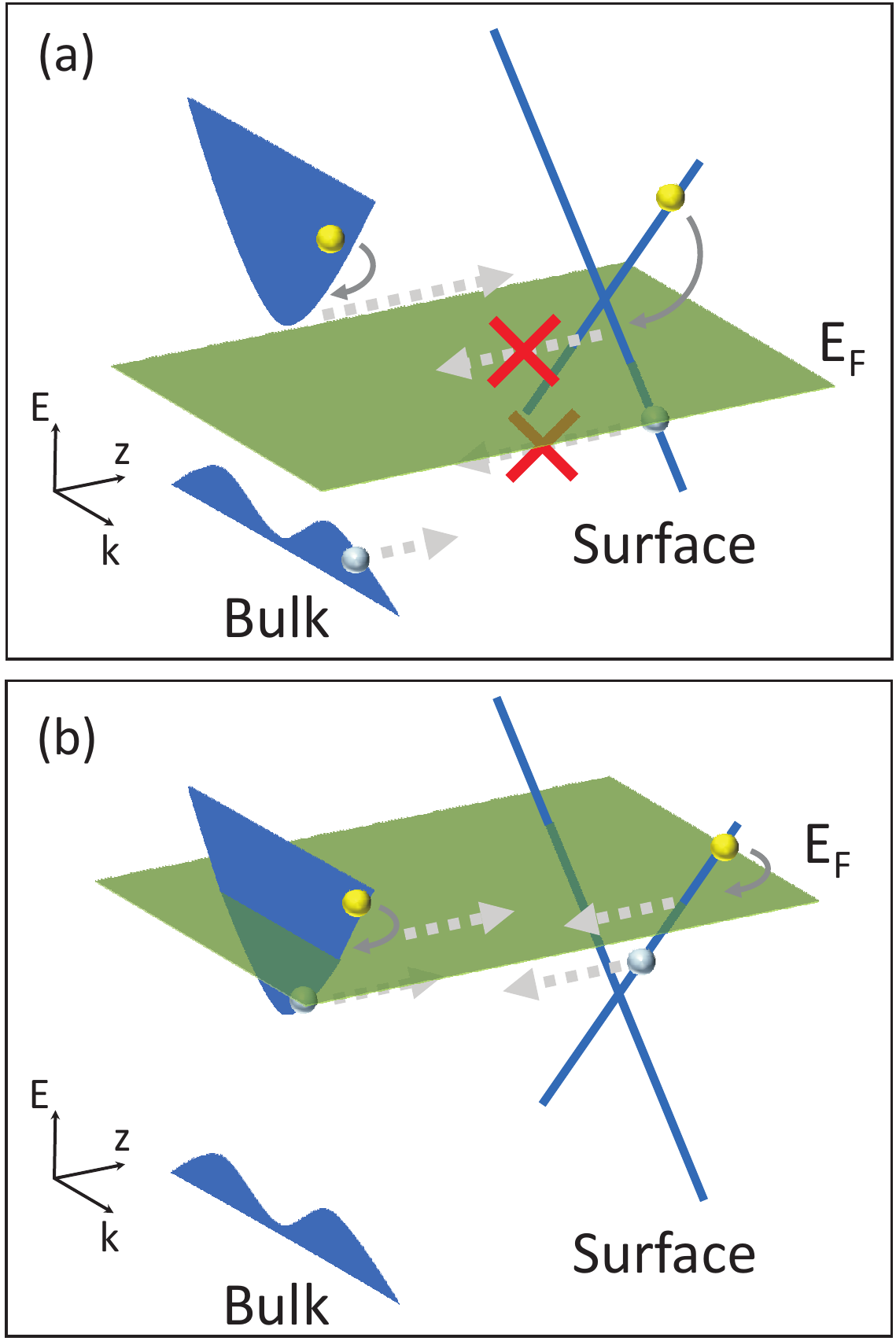}
\caption{Schematic representation of the microscopic process leading to the formation of a SPV. (a) In the bulk-insulating regime, carriers from the bulk diffuse towards the surface, while carriers from the surface decay quickly and are forbidden from diffusing into the bulk; bulk electrons diffuse more than holes leading to a negative charging of the surface. (b) In the bulk-conducting regime, carrier diffusion from the bulk is compensated by diffusion from the surface.}
\label{Fig4}
\end{figure}

To further verify this scenario, in Fig.~3(b) we show EDCs measured for $x$=0.55 without (red) and with laser pump at a pump-probe delay of $\Delta$t=-2 ps (blue). A quasi-static contribution to the SPV is clearly observed as an energy shift of $\sim$45 meV in the photoemission signal. The EDCs are extracted at \kpara=0 from the corresponding energy-momentum dispersions shown in the inset. We emphasize that the signal can be clearly observed at negative delays because it persists even at room temperature over time scales of $\sim$6 $\mu$s, and thus longer than previously observed for Bi$_2$Te$_2$Se. \cite{Neupane-PRL-2015} Such a long time scale corresponds to the maximum time separation between two consecutive pump pulses available in the present experiments.

Having observed a SPV in the intrinsic TI regime, we would like to further discuss the microscopic origin underlying this effect as well as its disappearance in the presence of bulk conduction. As shown in Fig.~4(a), after the femtosecond-laser excitation of a bulk-insulating TI, excited carriers at the surface experience a quick thermalization (within few ps), which brings them close to the Fermi energy. These carriers cannot diffuse into the bulk, since their energy falls within the bulk band gap. On the other hand the excited carriers within the bulk accumulate at the bottom of the BCB and at the top of the BVB so that electrons (yellow) and holes (gray) can diffuse towards the surface. However, electrons have lower mass compared to holes. Therefore electrons undergo a more efficient diffusion than holes (a similar effect, driven by the asymmetry between spin channels, gives rise to ultrafast spin diffusion \cite{Battiato-PRL-2010,Battiato-PRL-2017}). This leads to the injection of negative charge from the bulk into the surface. A bulk conducting TI has instead markedly different dynamics [see Fig.~4(b)]. After the initial thermalization, carriers both at the surface and in the bulk can diffuse. The difference in the transport properties in the bulk and at the surface can still lead to a SPV, but it will be much smaller than for a bulk-insulating TI (i.e., below our experimental sensitivity), as carrier diffusion from the bulk is compensated by diffusion from the surface. 

To summarize, by systematic tuning of the Sb concentration in the ternary TI (Bi$_{1-x}$Sb$_{x}$)$_2$Te$_3$, we have demonstrated the conditions under which a persistent SPV can be optically generated in this system at room temperature. We have examined the ultrafast response of excited electrons to femtosecond-laser excitation and disentangled the bulk from the surface contributions in the whole doping range. Finally, we have provided a microscopic picture of the underlying mechanisms responsible for the appearance of the SPV on ultrafast time scales. Our findings offer a potential platform for exploiting doping within a single TI compound in future optical applications, for example to print circuits that can be photo-activated on fs time scales while retaining the topological properties.

Financial support from the Deutsche Forschungsgemeinschaft (Grant No. SPP 1666) and the Impuls-und Vernetzungsfonds der Helmholtz-Gemeinschaft (Grant No. HRJRG-408) is gratefully acknowledged. M. B. acknowledges support from the Austrian Science Fund (FWF) through Lise Meitner grant M1925-N28.


\end{document}